\newtheorem{thrm}{Theorem}
\newtheorem{prep}{Proposition}
\newcommand{\mb}[1]{\color{black}#1}
\newcommand{\sa}[1]{\color{black}#1}

\documentclass[9pt,journal]{IEEEtran}
\usepackage{amssymb}
\usepackage{pgf,tikz}
\usepackage{mathrsfs}
\usetikzlibrary{arrows}
\usepackage{blindtext}
\usepackage{graphicx}
\usepackage{tkz-graph}
\usepackage{amsmath}
\newcommand{\mz}{\color{black}}

\newcommand{\mzr}{\color{black}}
\newcommand{\mkm}{\color{black}}
\newcommand{\mkmm}{\color{black}}
\newcommand{\mzrr}{\color{black}}
\newcommand{\mkmr}{\color{black}}
\newcommand{\mkmrr}{\color{black}}
\newcommand{\mkmb}{\color{black}}
\newcommand{\mkmbb}{\color{black}}

\newcommand{\mzrg}{\color{black}}

\newcommand{\blue}{\color{black}}

\newcommand{\mzfinal}{\color{black}}

\newcommand{\rep}{\eqref}
\newtheorem{remark}{\rm\bfseries Remark}[section]
\newtheorem{definition}{\rm\bfseries Definition}[section]

\newtheorem{lemma}{\rm\bfseries Lemma}[section]

\newtheorem{example}{\rm\bfseries Example}[section]

\GraphInit[vstyle = Normal]
\tikzset{
	LabelStyle/.style = { rectangle, rounded corners, draw,
		minimum width = 2em, fill = white!50,
		text = black, font = \bfseries },
	VertexStyle/.append style = { inner sep=1.5pt,
		font = \Large\bfseries},
}

\begin{document}
\title{Structural Controllability of {\blue a Consensus Network} with Multiple Leaders}
\author{\IEEEauthorblockN{Milad Kazemi M., Mohsen Zamani, and Zhiyong Chen,~\IEEEmembership{Senior Member,~IEEE}}
\thanks{M. Kazemi M. is with the Department of Mechanical Engineering, Amirkabir University of Technology (Tehran Polytechnic), Tehran, Iran.\newline kazemi.m@aut.ac.ir\par
M. Zamani and Z. Chen are with the School of Electrical Engineering and Computing, University of Newcastle, NSW 2308, Australia.\newline
mohsen.zamani@newcastle.edu.au, 
zhiyong.chen@newcastle.edu.au}}



\maketitle
{\sa
\begin{abstract}
This paper examines the structural controllability for a group of agents, called \textit{followers}, connected to each other {\mzr based on the } consensus law under commands of multiple \textit{leaders}, {\mzr which are agents with superior capabilities, over a} fixed communication topology. {\blue It is proved that the graph-theoretic sufficient and necessary condition for the set of followers to be structurally controllable under   the leaders' commands is leader-follower connectivity of the associated graph topology. This shrinks to graph connectivity for the case of solo leader. In the approach, we explicitly put into account the dependence among the entries of the system matrices for a consensus network using the  linear parameterization technique introduced in {\mzfinal \cite{morse2017Structural}}}.

\end{abstract}
}


%
\IEEEpeerreviewmaketitle
\section{Introduction}
In recent years, due to the importance of analyzing the complex systems, the notion of structural controllability {\mzrg has been} retaken into consideration. Defined as controllability of systems for almost every parameter values, structural controllability has a wide range of applications from {\mkmbb robotics \cite{Bowden2012} to biological systems} \cite{Vinayagam4976}. 
{\mb Lin \cite{lin1974structural} first introduced the structural controllability for single input {\sa linear time-invariant (LTI)} dynamical systems. He provided a {\sa graph-theoretic} representation that guarantees the {\sa structural} controllability {\mzr for} {\sa LTI systems, i.e. controllability}  for almost every parameter values. {\sa The new notion of controllability that Lin introduced encouraged other researchers to investigate the interaction among systems' parameters.} 

{\sa The authors of} \cite{shields1976structural} presented an algebraic representation of Lin's theorem and also extended the theorem to scrutinize the {\sa structural controllability}  {\sa for} multi-input {\sa LTI} systems. {\mkmbb The aforementioned studies dealt with dynamical systems in which each of the nodes represents  a first-order dynamical system. The structural controllability of  multi-input/multi-output (MIMO) high-order systems was investigated in \cite{Menichetti2016}, \cite{Chapman2014}, and \cite{CARVALHO2017123}.}     {\sa In most cases, one may need to examine beyond the fact that whether a system is structurally controllable or not}. For instance, when dealing with uncontrollable systems, declaring the maximum controllable subspace enables us to know our ability to control the system {\mzr (see e.g. \cite{hosoe1980determination},  \cite{poljak1990generic}, {\mkmbb \cite{liu2012control}, and \cite{Mousavi2017}})}. 

Moreover, in applications such as systems {\mzr biology,} the choice of input nodes (driver nodes) is so broad that selecting a proper set of nodes to ensure the controllability becomes a crucial problem in cell reprogramming or in cancer treatment {\mkmbb (see e.g. \cite{liu2011controllability}, \cite{Monshizadeh2014}, \cite{Pequito2016}, and \cite{Moothedath2018}).} {\sa The reference} \cite{liu2011controllability} determined the minimum number of driver nodes to guarantee the structural controllability of LTI  systems. {\sa In there, the authors} provided a polynomial algorithm to determine the driver nodes.  
}

These studies have a common assumption that the {\mkmrr nonzero} {\sa entries} of the pair $(A, B)$ are {\blue independent from each other
with free choices.}
Despite the wide application of this theory, it cannot analyze {\blue a system} with {\sa {\mzr the same} scalar values appearing} in more than one {\mzrg place}  in the pair $(A, B)$, {\sa which is indeed the case for many dynamical systems, i.e., entries of the pair $(A,B)$ cannot be assigned arbitrarily}. To overcome this dilemma, one can represent {\blue the system by a linearly} parameterized model. The authors of  \cite{morse2017Structural} extended the notion of structural controllability to {\blue linearly} parameterized systems with binary assumptions. This extension has a great application {\sa in the areas such as cooperative control {\mkmrr of} multi-agent systems where there exists 
{\blue inherent dependence} between the entries of the state and input matrices that capture the network topology.}


 In this {\mzr paper,} we examine the structural controllability for {\mzrg a} group of agents 
{\blue equipped with a consensus law}.   {\mzr The notion of controllability for such a setup} was first proposed by \cite{tanner2004controllability}. {\sa This reference} exploited controllability to examine the possibility that  a group of interconnected {\mzr agents} {\sa through consensus law}   can be steered to any desired configuration under the command of a single leader. {\sa Several} necessary and sufficient algebraic conditions {\sa for} controllability {\sa of multi-agent systems} based on eigenvectors of {\sa the associated} Laplacian graph {\sa were introduced in \cite{tanner2004controllability}}. 
 The problem was then developed {\sa further} in \cite{ji2006leader}, \cite{ji2007graph}, \cite{rahmani2006controlled}, and \cite{ji2008graph}. For instance, in \cite{ji2007graph}, it was concluded that devoid of eigenvalue sharing between the Laplacian matrix {\sa associated with} the follower set {\sa only} and the Laplacian matrix {\sa corresponding} {\mkmrr to} the whole topology is both necessary and sufficient for controllability. 
{\blue The reference \cite{ji2008graph} provided a necessary only condition for the controllability of followers under multiple leaders.}
 The {\sa graph-theoretic} representation of these {\sa results} {\sa was} introduced in \cite{zamani2009structural}. Moreover, the structural controllability of multi-agent systems with {\blue a} switching topology and the structural controllability of higher order multi-agent systems were studied in \cite{liu2013graph} and \cite{partovi2010structural}, respectively. Another sphere of research relies on the behavior of the system before {\sa and} after {\sa establishing link} or agent removal. Robustness of structural controllability against node and link removal {\sa was investigated} in \cite{jafari2010structural}  and \cite{rahimian2013structural}.  
 
{\sa In this paper, we exploit {\blue the} linear parameterization {\mzr technique} to deal with {\blue the dependance among the} entries of the pair $(A,B)$ when analyzing the structural controllability of interconnected linear systems. The reference \cite{zamani2009structural} addressed the same problem for the case of solo leader. Even though the results reported in \cite{zamani2009structural} {\mzrg are} correct, the authors neglected the above-mentioned {\blue inherent dependence} in the main proof stated there. 
{\mkmm Moreover, {\mzrr the authors in  \cite{liu2013graph} addressed the structural controllability for a group of interconnected agents with multiple leaders under a switching topology {\blue and provided the sufficient and necessary condition.}  Similar to  \cite{zamani2009structural},  in this reference there exists an implicit assumption about independence of entries appearing in {\blue the} $A$ and $B$ matrices. 
{\blue There seems no clear clue to fix the flaw of the proofs within the same framework. Therefore, we aim to 
provide an alternative rigorous proof using the linear parameterization technique recently developed in \cite{morse2017Structural}.
}

The rest of the paper is organized as follows. The terminology and concepts used in this paper are defined in Section \ref{sec: PRELIMINARIES}. The {\sa problem formulation is given} in Section \ref{sec: Problem Formulation}. {\sa We study the case where there exists only one leader among agents in Section \ref{sec: Structural Controllability of Single-input Systems}. Then the results of this section are exploited in Section \ref{sec: Structural controllability of multi-input systems} to examine the {\mzr multiple leaders case}. {\mzr Finally,} Section \ref{sec: Conclusion} concludes the paper.}
\section{PRELIMINARIES
\label{sec: PRELIMINARIES}}
\subsection{Structural Controllability}
Roughly speaking, the concept of controllability, as a paramount property of control systems, {\mz examines} the capability of a system to steer from  {\mz any} initial state to {\mz some desired} final {\mzr value} {\mz within} its entire configuration space {\mz under {\mzr a proper control law}. {\mz The answer to this examination}  is given by controllability tests like {\mzr the} {\mz Kalman's}  rank {\mz condition \cite{kalman1963mathematical} {\mkm or Popov-Belevitch-Hautus (PBH) controllability test \cite{hautus1969controllability}}.}  However, {\mz f}or many dynamical systems, the system's parameters are not precisely known and in some cases, the existence or {\mz absence  of system parameters} is the only accessible information. {\mz In addition to this}, some systems have time-variant parameters, and it is computationally hard to determine the controllability {\mz of these systems} during the {\mz whole} process. 

In order to overcome these {\mz challanges, Lin in his seminal paper  \cite{lin1974structural}} introduced the notion of structural controllability.
The Lin's theorem provided a test for checking the controllability of structured {\mz LTI systems,  which are LTI systems whose {\sa entries} of $A$ and $B$, i.e., {\blue the state and input matrices} are either zero or {\blue independently} free parameters.}  As defined {\mz   {\mzr in} \cite{lin1974structural}, the  pair of matrices
$(A, B)$ with each entry either  being a zero value or {\blue an arbitrarily chosen scalar not depending on other entries,}
is structurally controllable {\blue if} there exists a real {\blue controllable pair, say $(\bar{A},\bar{B})$,} with the same {\blue structure} of zero entries
as $(A, B)$. Consequently, the system is concluded to be controllable for almost every parameter values. 

{\blue The result in \cite{lin1974structural} is insightful but the concept of structural controllability {\mzfinal introduced in \cite{lin1974structural}} does not apply to 
systems with the inherent dependence among the  entries of $A$ and $B$.}  One should note that it is ubiquitous in many practical  scenarios like in biological systems, {\mz where one explores}  gene-gene interaction,  to have some of the interconnecting links be related to each other. 
{\blue To accommodate these systems, one needs to modify {\mzfinal the structural controllability definition in \cite{lin1974structural}} 
and propose new test tools. One 
way is to represent the system in the linearly parameterized form \cite{corfmat1976structurally}.} This approach is a convenient method to analyze LTI systems with parameter repetition in  {\mzfinal their} state and/or input matrices. {\mz We breifly review linear parameterization of structured systems in the next subsection.}

\subsection{Linear Parameterization}
Consider the LTI system   given as 
\begin{equation}
\dot{x}={\mkmrr A(w)}x+{\mkmrr B(w)}u,
\label{eq:lplti-formula}
\end{equation}
where ${\mz {\mkmrr A(w)} \in \mathbb R^{n \times n} }$ and ${\mz {\mkmrr B(w)} \in \mathbb R^{n \times m} }$  are functions of
an arbitrarily selected vector
 {\blue $  w= \begin{bmatrix}
 w_1&  w_2 & \ldots& w_\sigma
\end{bmatrix}^\top$.
 Suppose the matrices $(A,B)$ have {\mzfinal$p$} nonzero entries.
The definition in \cite{lin1974structural} only applies to the case that these {\mzfinal$p$} entries are exactly represented by $w$ with {\mzfinal$p=\sigma$}.

For the more general scenario with {\mzfinal$\sigma \leq p$,} 
the matrix pair $(A,B)$ can be linearly parameterized as}
\begin{equation}
\begin{array}{cc}
A_{n\times n}(w)=\sum_{k\in \mathbf{q}} c_k{\mkmrr w_k}r_{k1},&B_{n\times m}(w)=\sum_{k\in \mathbf{q}} c_k{\mkmrr w_k}r_{k2},
\end{array}
\label{eq:lplti-lp}
\end{equation} 
{\blue where  $\mathbf{q}=\{1,\ldots,\sigma\}$,  $c_k\in \mathbb{R}^n$, $r_{k1}\in \mathbb{R}^{1\times n}$, and $r_{k2}\in \mathbb{R}^{1\times m}$}. {\mz We provide the following example for further explanation of linear parameterization.}

\begin{example}\label{ex-formula-itself}
	Consider the following equation
	\begin{equation}\label{ex1-formula}
	\begin{array}{cl}
	\left[
	\begin{array}{c}
	\dot{v}_1\\
	\dot{v}_2\\
	\dot{v}_3
	\end{array}
	\right]
	&=
	\left[
	\begin{array}{ccc}
	-{\mkmrr w_1}-{\mkmrr w_2}-{\mkmrr w_3}&{\mkmrr w_2}&{\mkmrr w_3}\\
	{\mkmrr w_2}&-{\mkmrr w_2}&0\\
	{\mkmrr w_3}&0&-{\mkmrr w_3}
	\end{array}
	\right]
	\left[
	\begin{array}{c}
	v_1\\
	v_2\\
	v_3
	\end{array}
	\right]
	\\
	&\\
	& +
	\left[
	\begin{array}{c}
	{\mkmrr w_1}\\
	0\\
	0
	\end{array}
	\right]u.
	\end{array}
	\end{equation}
	The   above   LTI system attains the pair $(A,B)$ which is  a function {\mz of} $\begin{bmatrix}
		{\mkmrr w_1} & {\mkmrr w_2} & {\mkmrr w_3}
		\end{bmatrix}^\top$, and   its  associated  linear parameterization can be represented as
	\begin{equation}
	\begin{array}{lll}
	c_1=\left[
	\begin{array}{c}
	1\\
	0\\
	0
	\end{array}
	\right],&r_{11}=\left[
	\begin{array}{ccc}
	-1&0&0
	\end{array}
	\right],&r_{12}=1,\\
	&&\\
	c_2=\left[
	\begin{array}{c}
	-1\\
	1\\
	0
	\end{array}
	\right],&r_{21}=\left[
	\begin{array}{ccc}
	1&-1&0
	\end{array}
	\right],&r_{22}=0,\\
	&&\\
	c_3=\left[
	\begin{array}{c}
	-1\\
	0\\
	1
	\end{array}
	\right],&r_{31}=\left[
	\begin{array}{ccc}
	1&0&-1
	\end{array}
	\right],&r_{32}=0.\\
	\end{array}
	\end{equation}
	It is obvious that the vectors   $c_1$, $c_2$, and $c_3$ are linearly independent of each other and $\sigma=3$.
\end{example}

{\blue
The pair $(A(w),B(w))$ is called structurally controllable if there exists a parameter vector $w \in \mathbb{R}^\sigma$ for which  
the pair $(A(w),B(w))$ is controllable \cite{morse2017Structural}. We adapt the same definition in this paper. 

 It is worhtwhile noting that, for {\mzfinal$p=\sigma$},  the definitions of structural controllability in \cite{lin1974structural} and \cite{morse2017Structural}  are identical and  the  structural controllability of  the system in \eqref{eq:lplti-formula}  can be studied by the
results in  \cite{lin1974structural}.
The structural controllability of the system \eqref{eq:lplti-formula} for {\mzfinal$\sigma=p$} is also explored in \cite{corfmat1976structurally} from 
an algebraic point of view. 
Despite the well-approved algebraic structural controllability conditions in \cite{corfmat1976structurally}}, in most cases,  the graph-theoretic perspective provides more insights regarding hidden relations that undergo between system's parameters.  In the next subsection, we give a short  review on some graph theory concepts exploited in this paper.
 }

\subsection{Graph Notation}
{\mz  The reference \cite{lin1974structural}  exploited weighted-digraphs to represent dynamical systems. This graph representation not only shows the existence of directed interactions, or links,  between {\blue the entries of} $A$ and $B$, but also reveals the strength of those links.  This way of demonstrating dynamical systems enabled the author of \cite{lin1974structural} to introduce graph-theoretic descriptions} for structural controllability of single input LTI systems.
{\mz In this paper, we deploy the flow graph representation to study the dynamical system \eqref{eq:lplti-formula} from graph-theoretic {\mz point of view.}} 

{\mz Consider the weighted {\mkmrr graph} $\mathcal G$  with its node set $V=\{v_1,v_2,\ldots,v_{N_V}\}$, edge set $E=\{e_1, e_2, \ldots, e_{N_E}\}$, and  weight set $W$ corresponded to each link $W=\{(e_1,{\mkmrr w_1}), (e_2,{\mkmrr w_2}), \ldots, (e_{N_E},w_{N_E})\}$.} Let $N_V$ and $N_E$ be the number of the nodes and the edges, respectively. Then graph representation of the dynamical system \eqref{eq:lplti-formula}, which is called the flow graph  denoted by $\mathcal F_{\mathcal G}$,  is a digraph. {\mzrg It} includes $n+m$ vertices $V=\{v_1, v_2, \ldots, v_{n+m}\}$, where the input {\mz nodes} take the last $m$ indices, i.e., $v_{n+1}, \ldots, v_{n+m}$.  {\mz  Moreover, if the $\{i,j\}$ entry of {\mzrg the matrix $[A, B]$}  is {\mkmrr nonzero}, there exists a link from $v_j$ to $v_i$.} Two nodes are {\mz called \textit {neighbors}} if there exists an edge that corresponds these two nodes, and if all of {\mzrg the} nodes are neighbors to each other, the graph is called a {\mz \textit{complete graph}}. A {\mz \textit {path}} is a set of edges that connect a set of  {\mz distinct} nodes. The {\mz digraph is called {\mzrg \textit {connected}} provided that there exists a {\mz bidirectional} path between every two different vertices.

 	The {\mz flow graph $\mathcal{F}_\mathcal{G}$} has a \textit{spanning forest} rooted at {\mz vertices {\mb $v_{n+1}$,$v_{n+2}$, \ldots,$v_{n+m}$} if for any other node of the graph, {\blue say $v_j \in\{v_1, v_2,\ldots, v_n\}$,}  there exists  a path from one of the root nodes ${\mkmrr v_i}\in\{v_{n+1},v_{n+2}, \ldots,v_{n+m}\}$ to $v_j$. }
 	
Finally, the Laplacian {\mz matrix associated with the {\mkmrr graph} $\mathcal G$ } is defined as
\begin{equation}
L_{(i,j)}=\left\{
\begin{array}{lc}
\sum_{i\neq j}w_{ij}& i=j,\\
-w_{ij}&i\in {\sa\textnormal{neighborhood of }j},\\
0&{\sa\textnormal{otherwise.}}
\end{array}
\right.
\label{eq:laplacian}
\end{equation}

 For the sake of clear explanation, we represent the flow graph of the system (\ref{ex1-formula}) in Fig.~\ref{ex1-flowgraph}. 
	\begin{figure}
		\centering
		\begin{tikzpicture}
		\SetGraphUnit{2}
		\Vertex[Math,L=v_4,x=4.5,y=0] {D}
		\Vertex[Math,L=v_1,x=7,y=0] {B}
		\Vertex[Math,L=v_2,x=9.5,y=0] {C}
		\Vertex[Math,L=v_3,x=7,y=-2.5] {E}

		\tikzset{EdgeStyle/.append style = {->} }
		\Edge[label = ${\mkmrr w_1}$](D)(B)
		\tikzset{EdgeStyle/.append style = {->, bend left} }
		\Edge[label = ${\mkmrr w_2}$](B)(C)
		\Edge[label = ${\mkmrr w_2}$](C)(B)
		\Edge[label = ${\mkmrr w_3}$](E)(B)
		\Edge[label = ${\mkmrr w_3}$](B)(E)
		\Loop[dist = 3cm, dir = EA, label = $-{\mkmrr w_1}-{\mkmrr w_2}-{\mkmrr w_3}$](B.north)
		\Loop[dist = 3cm, dir = EA, label = $-{\mkmrr w_2}$](C.north)
		\Loop[dist = 3cm, dir = SO, label = $-{\mkmrr w_3}$](E.east)
		\tikzset{EdgeStyle/.append style = {bend left = 50}}
		
		\end{tikzpicture}\vspace{-17mm}
		\caption{The flow graph of   the system defined in (\ref{ex1-formula}).}
		\label{ex1-flowgraph}
	\end{figure}
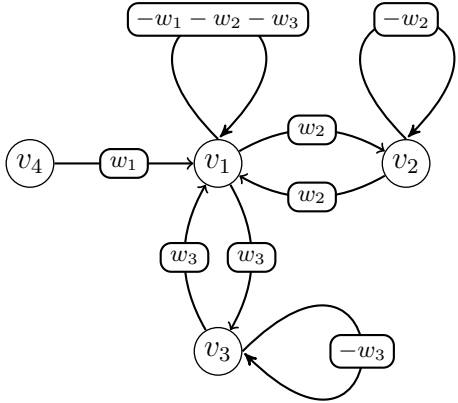
{\mkmr
\begin{remark}\label{rem1}
	It is {\mkmrr noteworthy that the vectors $c_i$ and $r_{i1}$ have   graph-theoretic implications in the system's associated flow graph. 
{\blue On one hand,} the nonzero entry of $c_i$, say $j$, captures an ingoing edge to node $v_j$ with weight $w_i$ in its associated flow graph. On the other hand, in $r_{i1}$ a nonzero entry expresses an outgoing edge from node $v_j$ in the associated flow graph {\mzrg and} $j$ is the index of that nonzero entry. This {\mzrg is} further demonstrated in the following example.}
	\end{remark}
\begin{example}
	Consider the following system
	\begin{equation}\label{ex2-formula}
	\dot{x}=\left[
	\begin{array}{ccc}
	{\mkmrr w_2}&0&{\mkmrr w_2}\\
	{\mkmrr w_1}&0&0\\
	{\mkmrr w_1}&0&0\\
	\end{array}
	\right]x.
	\end{equation}
	{\mkmrr One can verify that }this system {\mkmrr can be linearly parameterized with vectors $r_{i1}$, $r_{i2}$ and $c_i$ related to weight $w_1$ and $w_2$  as}
	\begin{equation}
	\begin{array}{lll}
	c_1=\left[
	\begin{array}{c}
	0\\
	1\\
	1
	\end{array}
	\right],&r_{11}=\left[
	\begin{array}{ccc}
	1&0&0
	\end{array}
	\right],&r_{12}=0,\\
	&&\\
	c_2=\left[
	\begin{array}{c}
	1\\
	0\\
	0
	\end{array}
	\right],&r_{21}=\left[
	\begin{array}{ccc}
	1&0&1
	\end{array}
	\right],&r_{22}=0.
	\end{array}
	\end{equation}
	The flow graph of this system is represented in Fig.~\ref{ex2-flowgraph}.
		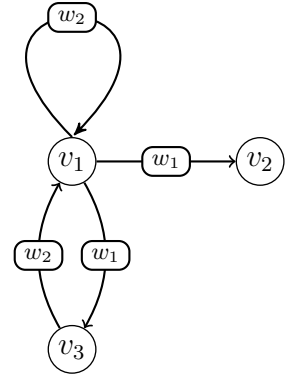
\begin{figure}
		\centering
		\begin{tikzpicture}
		\SetGraphUnit{2}
		\Vertex[Math,L=v_1,x=7,y=0] {B}
		\Vertex[Math,L=v_2,x=9.5,y=0] {C}
		\Vertex[Math,L=v_3,x=7,y=-2.5] {E}

		\tikzset{EdgeStyle/.append style = {->} }
		\Edge[label = ${\mkmrr w_1}$](B)(C)
		\tikzset{EdgeStyle/.append style = {->, bend left} }
		\Edge[label = ${\mkmrr w_1}$](B)(E)
		\Edge[label = ${\mkmrr w_2}$](E)(B)
		\Loop[dist = 3cm, dir = EA, label = ${\mkmrr w_2}$](B.north)
		
		\end{tikzpicture}
		\caption{The flow graph of the system defined in (\ref{ex2-formula}).}
		\label{ex2-flowgraph}
	\end{figure}

As stated in Remark \ref{rem1}, each of the nodes $v_2$ and $v_3$ in the corresponding flow graph (the indices of nonzero entries of $c_1$) has {\mkmrr an} ingoing edge with weight ${\mkmrr w_1}$. Moreover, the nonzero {\mkmrr entry} of $r_{11}$ {\mkmrr (the first entry)} determine{\mkmrr s} the outgoing edge from node {\mkmrr $v_1$ within the associated} flow graph. This system has {\mkmrr two outgoing edges} from $v_1$ {\mkmrr (one to node $v_2$ and the other one to node $v_3$)} with weight ${\mkmrr w_1}$.  It is worth noting that {\blue the} vectors $c_2$ and $r_{21}$ both have a nonzero first entry. This means that the node $v_1$ has a self loop with weight ${\mkmrr w_2}$.       
\end{example}		
}

\subsection{Matrix-Algebraic Terminology}
{\mz In this subsection, we state some notions and results that help us in establishing the main result of the  paper. }

{\mz The generic rank denoted by {\mzrg  $\textnormal{g-rank}[\cdot]$} of  linearly parameterized matrix $$M({\mkmrr w})=\begin{bmatrix}
{\mkmrr A(w)} & {\mkmrr B(w)}
\end{bmatrix}=\sum_{k\in \mathbf{q}} c_k {\mkmrr w_k}r_{k},
$$
where $r_{k}=\left[\begin{array}{cc}
r_{k1}&r_{k2}
\end{array}
\right]$,
  is the maximum rank of $M({\mkmrr w})$ for all possible values of ${\mkmrr w}$. Furthermore,}  the pair $(A, B)$ is irreducible if there exists no permutation matrix $Q$ such that
\begin{equation}
    \begin{array}{cc}
        QAQ^{-1}=
            \left[
            \begin{array}{cc}
                A_{11} & \mathbf{0} \\
                A_{12} & A_{22} 
            \end{array}
            \right],
        & 
        QB=
            \left[
            \begin{array}{c}
                \mathbf{0} \\
                B_2 
            \end{array}
            \right],
    \end{array}
\end{equation}
where $A_{11}\in \mathbb{R}^{h\times h}$, $A_{12}\in \mathbb{R}^{(n-h)\times h}$, $A_{22}\in \mathbb{R}^{(n-h)\times (n-h)}$, {\mkm $B_{2}\in \mathbb{R}^{(n-h)\times m}$}, and $1\leq h\leq n$.
It  then becomes evident that the system is structurally controllable if the pair $\left({\mkmrr A}, {\mkmrr B}\right)$ is irreducible 
 and its associated g-rank is equal to the number of states, i.e., $n$.   	

The irreducibility of the system has a {\sa graph-theoretic} implication which is stated in the following proposition.

\begin{prep}
	\cite{mayeda1981structural}
	  The pair $\left({\mkmrr A}, {\mkmrr B}\right)$ is irreducible if and only if the assocaited flow graph $\mathcal{F}_\mathcal{G}$ has a spanning forest rooted at 
	 $v_{n+1}$, $\ldots$, $v_{n+m}$.
	\label{p1}
\end{prep}
{\mkmbb
\begin{remark}
Proposition \ref{p1} was initially developed to address the matrix pairs satisfying the unitary assumption which means that each weight appears only in one entry of the matrix pair $(A(w), B(w))$, i.e., $\sigma=l$; however, the same proof applies to  the case in  which $\sigma\le l$ without any modification \cite{morse2017Structural}.
\end{remark}
}

\section{Problem Formulation
\label{sec: Problem Formulation}}

{
Our goal in this paper is to investigate structural controllability for {\mzrg a} group of interconnected systems with fixed topology {\mzrg of no self-loops}. We consider {\mzr a group of $N$  interconnected agents  and  focus} on the leader-follower framework, where there exist $l$ agents with superior capabilities and access to external commands which we refer to as leaders, while the remainder of agents take the follower role. 
%
 Without  loss of generality, the last $l$ agents are considered as leaders manipulated by some external input, and  the remaining $N-l$ agents are  controlled by the {\mzr consensus law}.

Each follower can be modeled as a point mass exerted  
by an external load as
\begin{equation} \label{eq:cons}
\dot{x}_i=-\sum_{j\in \mathcal{N}_i}w_{ij}(x_i-x_j),
\end{equation}
where $\mathcal{N}_i$ is the set that captures the neighbors of the agent $i$, and $w_{ij}\neq 0$
is weight
of the edge from $j$ to $i$.
{\mkmrr For the sake of simplicity in the notation we suppose that there exists a bijective mapping between two sets $\{\begin{array}{c|c}
	w_{ij}&i\in \mathcal N_j,i<j
	\end{array}\}$ and $\{w_1, w_2, \ldots, w_{\alpha}\}$. In the rest of this paper, we exploit $w_k$ instead of $w_{ij}$.}

\begin{example} {\blue
The system \eqref{ex1-formula}	 can be rewritten as 
\begin{equation} 
	\left[
	\begin{array}{c}
	\dot{v}_1\\
	\dot{v}_2\\
	\dot{v}_3
	\end{array}
	\right]
	=
	- \left[
	\begin{array}{ccc} w_1(v_1-u) +w_2(v_1-v_2) + w_3(v_1-v_3) \\
w_2 (v_2 - v_1) \\
w_3(v_3-v_1) 
	\end{array}
	\right], 
	\end{equation}
so, the graph topology for the consensus network, called the communication topology, is shown in Fig.~\ref{graphExample}, 
while the flow graph is in Fig.~\ref{ex1-flowgraph}.  }
 
\end{example} 	
	
 The leaders do not follow the {\mzr law in \eqref{eq:cons}}, and are controlled exclusively by {\mzr some external input expressed as}
\begin{equation}
\dot{x}_j=u_j^\star,
\end{equation}
where $j$ defines the index number of leader vertices  $j\in\{N-l+1,\ldots,N\}$.  The aggregated dynamical model of the whole interconnected system can be obtained as \cite{zamani2009structural}
{\mkmbb
{\mzrg\begin{equation} \label{eq:agg}
\dot{\bar x}=\left[\begin{array}{cc}
A_{(N-l)\times(N-l)}&B_{(N-l)\times l}\\
0_{l\times(N-l)}&0_{l\times l}
\end{array}\right]\bar x+
\left[\begin{array}{c}
0_{(N-l)\times 1}\\
u_{l\times 1}^\star
\end{array}\right].
\end{equation}}
}
{\mkmbb
{\mzrg In the set of equations in \eqref{eq:agg} the leaders' positions can be seen as inputs to autonomous dynamics captured by followers only.  Then the part of above dynamics only associated with followers can be simplified as}} 
\begin{equation}
\dot{x}=Ax+Bu,
\label{system}
\end{equation}
{\mkmbb
{\mzrg where $A=-L_{ff}$, where $L_{ff}$ is the part of  Laplacian matrix} {\mzr associated with followers only}. {\mzrg And, the matrix $B$ only captures the interactions between followers and leaders.}} 
Our task in paper is to explore the controllability of the system \eqref{system} under the commands of  multiple leaders and establish a  {\sa graph-theoretic} condition which is both sufficient and necessary  for  guaranteeing  structural controllability.

To explore the controllability of multi-agent systems, in the following sections, we first consider the case of single leader  and derive the necessary and sufficient condition {\mzr for this setup. We then extend the theorem to the case with more than one leader.} 
\section{Structural Controllability of {\sa multi-agent Systems with single leader}
	\label{sec: Structural Controllability of Single-input Systems}}
 In this section, a {\mzr  sufficient and  necessary} condition for structural controllability of a group of agents under a solo leader with fixed communication topology is introduced. 
 {\mzr To this end, let us first consider an edge {\mkmbb with the weight $w_k$} that {\mzrg connects} two vertices $v_i$ and $v_{j}$ within the {\mzfinal flow} graph $\mathcal F_\mathcal G$ that captures the interactions between entries of $A$ and $B$ matrices in \eqref{system}.} Without loss of generality, we assume that $i<j$. {\mzr Then these two vectors  {\mkmbb $c_k\in\mathbb{R}^{n\times1}$, and $r_{k1}\in\mathbb{R}^{1\times n }$} have zero entries except their $i$ and $j$ entries i.e.}
{\mkmbb
\begin{equation}\label{eq:cere}
\begin{array}{cc}
c_k^{(i)}=-1,&c_k^{(j)}=1,\\
&\\
r_{k1}^{(i)}=1,&r_{k1}^{(j)}=-1.
\end{array}
\end{equation}
}
Let us {\mzr now} introduce the set {\mkmbb $\mathbf{s}=\{i_1, \ldots, {\mkmb{i_s}}\}{\sa \subset} \mathbf{q}$ 
where   $s$ is the cardinality of set $\mathbf{s}$. {\mzr Then the}  matrices $C_\mathbf{s}$, $R_\mathbf{s}$, and ${\mkmrr W_\mathbf{s}}$ {\mzr can be} defined as{\mzr
\begin{equation}
\begin{array}{l}
C_\mathbf{s}\triangleq\left[
\begin{array}{cccc}
c_{i_{_{1}}}&c_{i_{_{2}}}&\ldots&c_{i_{_s}}
\end{array}
\right],
\\
\\
R_\mathbf{s}\triangleq\left[
\begin{array}{cccc}
r_{i_{_{1}}}^\top&r_{i_{_{2}}}^\top&\ldots&r_{i_{_s}}^\top
\end{array}
\right]^\top,
\\
\\ 
{\mkmrr W_\mathbf{s}\triangleq\textnormal{diag}\left[
\begin{array}{cccc}
w_{i_{_1}}&w_{i_{_2}}&\ldots&w_{i_{_s}}
\end{array}
\right].}
\end{array}
\end{equation} }
{\sa
\begin{example}
	Consider the graph topology of system represented in {\mzr \eqref{ex1-formula}} under the leadership of node $v_4$. This system has, as {\mzr discussed in Example \ref{ex-formula-itself},} three independent {\mzr parameters} i.e. {\mzr $\sigma=3$}. {\mzfinal Thus,} {\mzr for this case}, we have
		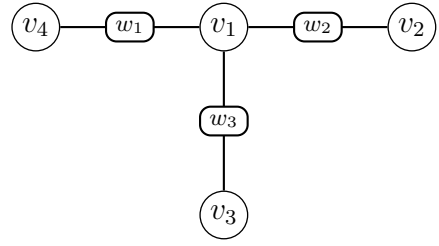
\begin{figure}
			\centering
		\begin{tikzpicture}
		\SetGraphUnit{2}
		\Vertex[Math,L=v_4,x=4.5,y=0] {D}
		\Vertex[Math,L=v_1,x=7,y=0] {B}
		\Vertex[Math,L=v_2,x=9.5,y=0] {C}
		\Vertex[Math,L=v_3,x=7,y=-2.5] {E}

		\tikzset{EdgeStyle/.append}
		\Edge[label = ${\mkmrr w_1}$](D)(B)
		\Edge[label = ${\mkmrr w_2}$](C)(B)
		\Edge[label = ${\mkmrr w_3}$](B)(E)
		\end{tikzpicture}
		\caption{The {\blue communication topology} associated with the system in \eqref{ex1-formula}.}
		\label{graphExample} 
	\end{figure}

{\mkmbb	
	\begin{equation}
	\begin{array}{ll}
	c_1=c_{v_1v_4}=\left[
	\begin{array}{c}
	1\\
	0\\
	0
	\end{array}
	\right],&r_1=r_{v_1v_4}=\left[
	\begin{array}{ccc|c}
	-1&0&0&1
	\end{array}
	\right],\\
	&\\
	c_2=c_{v_1v_2}=\left[
	\begin{array}{c}
	-1\\
	1\\
	0
	\end{array}
	\right],&r_2=r_{v_1v_2}=\left[
	\begin{array}{ccc|c}
	1&-1&0&0
	\end{array}
	\right],\\
	&\\
	c_3=c_{v_1v_3}=\left[
	\begin{array}{c}
	-1\\
	0\\
	1
	\end{array}
	\right],&r_3=r_{v_1v_3}=\left[
	\begin{array}{ccc|c}
	1&0&-1&0
	\end{array}
	\right].\\
	\end{array}
	\end{equation}
	}
Hence, if we consider $\mathbf{s}=\mathbf{q}$ the $C_\mathbf{s}$, $R_\mathbf{s}$ and ${\mkmrr W_\mathbf{s}}$ can be written {\mzrg  as} follows	
	\begin{equation}
\begin{array}{c}

C_{\mathbf{s}}=\left[
\begin{array}{ccc}
1&-1&-1\\
0&1&0\\
0&0&1
\end{array}
\right],
\\
\\
R_{\mathbf{s}}=\left[
\begin{array}{ccc|c}
-1&0&0&1\\
1&-1&0&0\\
1&0&-1&0
\end{array}
\right],
\\
\\
W_{\mathbf{s}}=\left[
\begin{array}{ccc}
{\mkmrr w_1}&0&0\\
0&{\mkmrr w_2}&0\\
0&0&{\mkmrr w_3}
\end{array}
\right].
\end{array}
\end{equation}

\end{example}

}
We need to introduce the notion of transfer matrix for establishing the main result of this paper. The transfer matrix of $\{(c_i,r_{i1},r_{i2})\vert i\in \mathbf{q}\}$, denoted by $T$, is
a block matrix defined as
\begin{equation}
T_{i,j}=
\left\{
\begin{array}{ll}
r_{i1}c_j&i,j\in\mathbf{q}\\
r_{i2}&i\in\mathbf{q},\; j={\mkmr \sigma+1}
\end{array}
\right. .
\label{eq:transgrph}
\end{equation}
We refer to the  graph  associated with the transfer matrix $T$ as transfer graph denoted by $\mathcal T$. This is a directed graph with {\mzr $\sigma+1$ vertices $\gamma_1, \ldots, \gamma_\sigma, {\mkmr \gamma_{\sigma+1}}$} and an {\sa edge} from {\sa node} $\gamma_j$ to $\gamma_i$ whenever ${\mkmrr T_{i,j}}$ is nonzero.

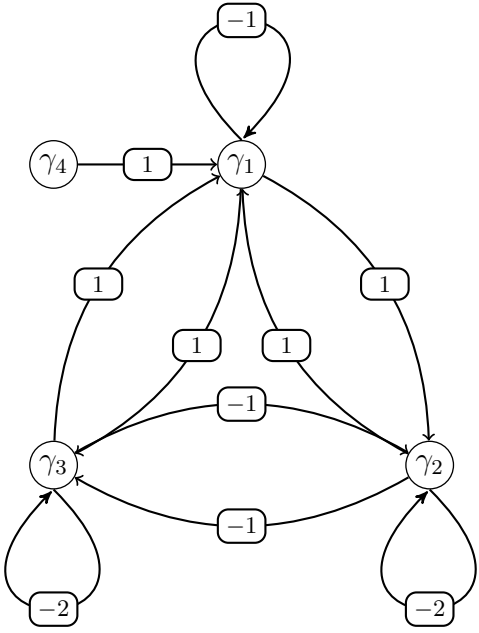
\begin{figure}
	\centering
	\begin{tikzpicture}
	\SetGraphUnit{2}
	\Vertex[Math,L=\gamma_{4},x=2.5,y=0] {A}
	\Vertex[Math,L=\gamma_1,x=5,y=0] {B}
	\Vertex[Math,L=\gamma_2,x=7.5,y=-4] {C}
	\Vertex[Math,L=\gamma_3,x=2.5,y=-4] {E}

	\tikzset{EdgeStyle/.append style = {->} }
	\Edge[label = $1$](A)(B)
	\tikzset{EdgeStyle/.append style = {->, bend left} }
	\Edge[label = $1$](C)(B)
	\Edge[label = $1$](B)(C)
	
	\Edge[label = $1$](B)(E)
	\Edge[label = $1$](E)(B)
	
	\Edge[label = $-1$](E)(C)
	\Edge[label = $-1$](C)(E)

	\Loop[dist = 3cm, dir = EA, label = $-1$](B.north)
	\Loop[dist = 3cm, dir = WE, label = $-2$](C.south)
	\Loop[dist = 3cm, dir = WE, label = $-2$](E.south)
	\tikzset{EdgeStyle/.append style = {bend left = 50}}
	\end{tikzpicture}
	\caption{The transfer graph of the system defined in (\ref{ex1-formula}).}
	\label{fig:ex1-transfergraph}
\end{figure}
\begin{example}
	{\mzr Consider} the system (\ref{ex1-formula}), if {\mzfinal  $\mathbf s=\{1\}\subset\mathbf{q}, C_\mathbf s, R_\mathbf s$, and ${\mkmrr W_\mathbf s}$} are 
	\begin{equation}
	\begin{array}{l}
	
	C_\mathbf{s}=\left[
	\begin{array}{c}
	1\\
	0\\
	0
	\end{array}
	\right],
	\\
	\\
	R_\mathbf{s}=\left[
	\begin{array}{ccc|c}
	-1&0&0&1\\
	\end{array}
	\right],
	\\
	\\
	{\mkmrr W_\mathbf{s}}=\left[
	\begin{array}{c}
	{\mkmrr w_1}
	\end{array}
	\right].
	\end{array}
	\end{equation}
	Similarly, $C_{\mathbf{q}-\mathbf{s}}, R_{\mathbf{q}-\mathbf{s}}$, and $W_{\mathbf{q}-\mathbf{s}}$ are
	\begin{equation}
	\begin{array}{c}
	
	C_{\mathbf{q}-\mathbf{s}}=\left[
	\begin{array}{cc}
	-1&-1\\
	1&0\\
	0&1
	\end{array}
	\right],
	\\
	\\
	R_{\mathbf{q}-\mathbf{s}}=\left[
	\begin{array}{ccc|c}
	1&-1&0&0\\
	1&0&-1&0
	\end{array}
	\right],
	\\
	\\
	W_{\mathbf{q}-\mathbf{s}}=\left[
	\begin{array}{cc}
	{\mkmrr w_2}&0\\
	0&{\mkmrr w_3}
	\end{array}
	\right].
	\end{array}
	\end{equation}
	
	{\mkmr For} the system (\ref{ex1-formula}) the transfer matrix $T$ can be respresented as
	\begin{equation}
	T=\left[
	\begin{array}{cccc}
	 -1 & 1 & 1 &1  \\
	 1 & -2 & -1 & 0  \\
	 1 & -1 & -2&0
	\end{array}
	\right].
	\end{equation}
The transfer graph of the above transfer matrix is shown in Fig.~\ref{fig:ex1-transfergraph}.
\end{example}
{\mkmr
\begin{remark}\label{rem2}
As defined in \eqref{eq:transgrph},  {\blue the entry $T_{i,j}$ with $i,j\in\mathbf{q}$} is obtained by inner product of two vectors ${\mkmrr r_{i1}}$ and $c_j$ and this means $r_{i1}c_j=\Sigma_{k=1}^{n}{\mkmrr r_{i1}^{(k)}}{\mkmrr c_{j}^{(k)}}={\mkmrr r_{i1}^{(1)}}{\mkmrr c_{j}^{(1)}}+\ldots+{\mkmrr r_{i1}^{(n)}}{\mkmrr c_{j}^{(n)}}$ where $ r_{i1}^{(k)}$ and $c_j^{(k)}$ represent  the $k$th entry of $r_{i1}$ and  $c_j$, accordingly. Each of these {\mzfinal terms} has a graph representation. The terms ${\mkmrr r_{i1}^{(k)}}$  and ${\mkmrr c_{j}^{(k)}}$ seek  for an outgoing edge from node $v_k$ with weight $w_i$, and another ingoing edge to node $v_k$ with weight $w_j$, respectively.
\end{remark}
}

 {\mzr Theorem \ref{mainresult}} represents a {\sa graph-theoretic} sufficient and necessary condition that guarantees the structural controllability among interconnected agents with fixed topology {\sa under single leader}. {\sa {\mzr Before we state this result}, we first need to introduce some results which {\mzr  enable} us to establish the main theorem of this section}. Proposition \ref{p1} and Lemma \ref{l2} {\mzr provide results in order to link the irreducibility of the system \eqref{system} to characteristics of its associated  transfer graph $\mathcal{T}$, which  is used in the proof of Theorem \ref{mainresult}.} 

{\mzrr We first introduce the following lemma which  is inspired by the result in }  \cite{morse2017Structural}.  {\mzrr In  \cite{morse2017Structural} the authors established a connection between the irreducibility property of  a linear parameterized representation  and the structure of its associated transfer graph when  just $1$ and $0$ values appear within the corresponding $c_i$,  $r_{i1}$ and $r_{i2}$ vectors. This assumption does not hold in our case; thus, we need to extend the proof initially stated {\mkmr in} \cite{morse2017Structural}.}
{\mkmm
\begin{lemma}
	{\mkmm If the pair $(A, B)$ for the system \eqref{system} is irreducible, then the associated transfer graph $\mathcal{T}$ has a spanning tree rooted at ${\mkmr \gamma_{\sigma+1}}$}.
	\label{l2}
\end{lemma}

{\mkmm  {\mzrr In order to prove the above lemma, we follow the approach of \cite{morse2017Structural}. To this end, we first} introduce the concept of {\mzrr \textit {line graph}}. The line graph  associated with the  directed graph {\mzrr ${\mkmrr\mathcal{F}_\mathcal{G}}$} is {\mzrr also a   directed graph ${\mkmrr \mathcal{L}_\mathcal{G}}$}  that represents the adjacencies between the edges of {\mzrr ${\mkmrr\mathcal{F}_\mathcal{G}}$}. {\mzrr Each edge of the original graph ${\mkmrr\mathcal{F}_\mathcal{G}}$ is presented by a node in its associated line graph ${\mkmrr \mathcal{L}_\mathcal{G}}$. Thus,  the number of edges in {\mzrr ${\mkmrr\mathcal{F}_\mathcal{G}}$} is equal to the number of veritices in the corresponding ${\mkmrr \mathcal{L}_\mathcal{G}}$. It is worth mentioning that two edges with the same start and end nodes but different {\mzrg weights}   in ${\mkmrr\mathcal{F}_\mathcal{G}}$, are  captured as  different nodes in ${\mkmrr \mathcal{L}_\mathcal{G}}$.}
 {\mkmr In order to construct ${\mathcal L}_{\mathcal G},$ one needs parameters associated with each edge in the flow graph ${\mkmrr\mathcal{F}_\mathcal{G}}$, namely start node, end node, and the corresponding weight. Then, the node $ijk$ is connected to the node $jj'k'$ in the line graph if there exist two edges in the flow graph; one from the node $v_i$ to the node $v_j$ with weight ${\mkmrr w_k}$ and the other one from the node $v_j$ to $v_{j'}$ with weight $w_{k'}$} 

\vspace{-4mm}
 {\mkmr \subsection*{Proof of Lemma \ref{l2}}
 		
 		Suppose that the relation between entries of $(A,B)$ is captured by the flow graph ${\mkmrr \mathcal{F}_\mathcal{G}}$ and 	
 		includes $\sigma$ independent parameters. Let us consider the equivalent class $H_o=\left\{\begin{array}{l|l}
 	ijk\in V_{\mathcal{L}}& k=o
 	\end{array}\right\}$ with $V_{\mathcal{L}}$ being the node set of ${\mkmrr \mathcal{L}_\mathcal{G}}$ and $o\in\{1,\ldots,\sigma\}$. 
 	 Hence, a quotient graph like $\widehat{\mathcal{L}}$ can be inaugurated in such way that   it has $\sigma$ nodes corresponding to its independent weights. This quotient graph has an edge between two nodes, if there exists at least one edge between the two sets of nodes in the line graph corresponding to these two nodes. In other words, the quotient graph has an edge from node $k$ to node $k'$, if there exists a node $v_j$ such that two edges exist in the flow graph: one from an arbitrary node $v_i$ to node $v_j$ with weight ${\mkmrr w_k}$ and the other one from $v_j$ to an arbitrary node $v_{j'}$ with weight $w_{k'}$. 

Now, we can {\mkmrr focus on} the transfer matrix.
Based on \eqref{eq:transgrph},  the transfer graph $\mathcal{T}$ has an edge from node $\gamma_k$ to $\gamma_{k'}$ if $r_{k1}c_{k'}\neq0$. 
Based on \rep{eq:cere}, we know that for all $k$'s at most two entries of $c_k$ and $r_{k1}$ are nonzero. 
{\blue In particular, 
 the entry $1$ of {\blue $c_k$}  has a higher index than that of $-1$}. The reverse holds for $r_{k1}$}. Therefore, if {\mkmrr one} calculates {\mkmrr $r_{k1}c_{k'}=\Sigma_{j=1}^{n}r_{k1}^{(j)}c_{k'}^{(j)}$}, {\mkmrr there exists at most two nonzero summand say $r_{k1}^{(j_1)}c_{k'}^{(j_1)}+r_{k1}^{(j_2)}c_{k'}^{(j_2)}$ which also have the same sign. Hence, one nonzero summands guarantees the existence of an edge from $\gamma_k$ to $\gamma_{k'}$ in the corresponding transfer graph.}
{\mkmrr This is analogous to } {\mzrg having} an edge from an arbitrary node $v_i$ to $v_j$ {\mkmrr($j\in\{j_1,j_2\}$)} with weight ${\mkmrr w_k}$ and another edge from $v_j$ to an arbitrary node $v_{j'}$ with weight $w_{k'}$.
{\mkmrr Let us now introduce the induced subgraph $\widehat{\mathcal{T}}$  which is obtained from $\mathcal{T}$ by deleting the node $\gamma_{\sigma+1}$ and its associated edges. }

Now based on {\mzfinal above-mentioned} definition{\mkmr s}, we can conclude that $\widehat{\mathcal{L}}$
and $\widehat{\mathcal{T}}$
are isomorphic with the bijection that maps vertex $i$ in $\widehat{\mathcal{L}}$
to
vertex $\gamma_i$ in $\widehat{\mathcal{T}}$. {\mkmrr If the pair  $(A, B)$ is irreducible, by Proposition \ref{p1}, ${\mkmrr \mathcal{F}_\mathcal{G}}$
has a spanning forest rooted at $v_{n+1}, v_{n+2}, \ldots$, and $v_{n+m}$. Thus, the associated $\widehat{\mathcal{L}}$ has a spanning forest rooted at $\{\begin{array}{c|c}
i&i\in\mathbf{q},r_{i2}\neq0
\end{array}\}$ that capture the weights of edges corresponding to nodes $v_{n+1}, v_{n+2}, \ldots$ , $v_{n+m}$ in the original flow graph. 
Consequently, $\widehat{\mathcal{T}}$
has a spanning forest rooted at {\mzfinal  $\gamma_i$s} {\mkmrr where {\mzfinal $i$s} are the indices of roots of spanning forest for the quotient graph}. Since there exist{\mkmrr s} an edge from $\gamma_{\sigma+1}$ to each of such $\gamma_i$'s in the transfer graph $\mathcal{T}${\mzrg, it } has a spanning tree rooted at ${\mkmr \gamma_{\sigma+1}}$.
}}

{\mzrg The following theorem states  the necessary and sufficient condition for structural controllability of linear parameterized systems.}
\begin{prep}
	\cite{morse2017Structural}	A linearly parameterized matrix pair $(A, B)$ is structurally controllable if and only if 
	\begin{equation}
\label{eq:rankcondition}	\min_{\mathbf{s} \subset \mathbf{q}}(\textnormal{rank}C_\mathbf{s}+ \textnormal{rank}R_{\mathbf{q}-\mathbf{s}})=n
	\end{equation} 
	and $\mathcal{T}$ has a spanning 
	tree rooted at ${\mkmr \gamma_{\sigma+1}}$.
	\label{p3}
\end{prep}


{\mzr We can now present the main result of this section. This theorem states that  the connectivity of the system is both the necessary and the sufficient condition for guaranteeing the structural controllability of the system \eqref{system} when there exists only one leader in the network.} 

\begin{thrm}
  Consider the  system (\ref{system}) under the {\sa fixed} communication topology $\mathcal{G}$ and {\blue a} single {\sa leader, $l=1$.} {\mzr This system is} structurally controllable if and only if $\mathcal{G}$ is connected.
	\label{mainresult}
\end{thrm}
\vspace{-3mm}

 \subsection*{Proof of {\blue Sufficiency}}
 {\sa Given the topology is \textit{connected}}, {\sa there exists} at least $N-1$ edges among these nodes. {\mkmb We first assume that the connected graph has exactly $N-1$ edges. In this case, the parameters {\blue $c_i$'s} corresponding to these $N-1$ edges are independent of each other, because each one of them establishes a link between two different vertices. The same holds for $\left[\begin{array}{cc}
     r_{i1}&r_{i2}
     \end{array}\right]$ vectors.
     } {\sa In the worst case {\mzr scenario}, there exists only one link from the leader to one of the followers}. 
     Thus, {\mzr a   connected  topology} consists  {\mzrg of }at least {\mzr $N-1=\sigma$} independent $c_i$ vectors. A similar {\mzr argument} can be applied {\mzr for counting the number of independent  $\left[\begin{array}{cc}
     r_{i1}&r_{i2}
     \end{array}\right]$ vectors}. {\sa Thus as the columns of the matrix $C_\mathbf{s}$ are linearly independent of each other, one can observe that $\textnormal{rank}C_\mathbf{s}=f$. {\mzr By exploiting the same argument, we can conclude that} {\mzr $\textnormal{rank}R_{\mathbf{q}-\mathbf{s}}=\sigma-f$}, where $\sigma$ and $f$ are the cardinality of the sets $\mathbf{q}$ and $\mathbf{s}$, respectively}. {\sa We now {\mzrg invoke} the} Proposition \ref{p3}. {\mkm It} {\sa  is easy to see that its first part} is satisfied {\mzr i.e. ($\min_{\mathbf{s} \subset \mathbf{q}}(\textnormal{rank}C_\mathbf{s}+ \textnormal{rank}R_{\mathbf{q}-\mathbf{s}})=N-1$)}. {\mkmb Now suppose that the number of weights exceeds the number of states, i.e., we have more than $N-1$ edges, it is obvious that the {\mzfinal  $c_i$ vectors} are not independent of each other anymore. The same holds for $\left[\begin{array}{cc}
     r_{i1}&r_{i2}
     \end{array}\right]$ vectors. Let us introduce a subgraph of this topology which is connected, has $N-1$ edges, and contains no simple cycles. Such a subgraph, as already established, satisfies the rank condition in (\ref{eq:rankcondition}) and one can easily conclude that the same should hold for the original graph.}
     

     
     {\sa On the other hand, if we have a connected topology with single leader, it is easy to see that there exists a spanning tree rooted at leader's node. Moreover, Proposition \ref{p1}  declares that the irreducibility is equivalent to existence of a spanning forest rooted at $v_{n+1}, \ldots, v_{n+m}$. Due to the fact that the system has only one leader, the notion of spanning forest can be considered analogous to the notion of spanning tree. Hence, the system is irreducible. 
     	Furthermore, based on Lemma \ref{l2}, we can  conclude that $\mathcal{T}$ has {\blue a} spanning tree rooted at ${\mkmr \gamma_{\sigma+1}}$. Hence, the connectivity of the topology guarantees the existence of {\mzrg a} spanning tree for transfer graph.}
     \vspace{-4mm}
     \subsection*{Proof of Necessity}
     {\sa We use proof by contradiction to establish this part}. {\sa Suppose that} the system {\sa was} structurally controllable, while it was not connected.
     {\mzr Then the  system could} be represented as}
     \begin{equation}
     \dot{x}=
     \left[
     \begin{array}{cc}
     L_{d_1\times d_1}&\mathbf{0}_{d_1\times d_2}\\
     \mathbf{0}_{d_2\times d_1}&L_{d_2\times d_2}
     \end{array}
     \right]x+
     \left[
     \begin{array}{c}
     b\\
     \mathbf{0}_{d_1+d_2-1}
     \end{array}
     \right]u.
     \label{system-contradiction}
     \end{equation}
The {\mzr above system can be considered as two separated {\mzr subsystems}: the connected topology which includes the leader and the rest of the topology. Based on this definition, $d_1$ represents the number of nodes   in the connected topology that includes the leader and the remaining $d_2$ nodes are considered as a different subsystem.} According to Kalman's theorem the controllability matrix for the {\mzr system (\ref{system-contradiction})}  
     can be obtained as
     \begin{equation}
     \left[
     \begin{array}{cccc}
     \bar{B}&\bar{A}\bar{B}&\ldots&{\sa \bar{A}^{n}}\bar{B}
     \end{array}
     \right]
     =
     \left[
     \begin{array}{llll}
     b&\star&\ldots&\star\\
     0&\star&&\star\\
     \vdots&\vdots&\ddots&\vdots\\
     0&\star&\ldots&\star\\
     \hline
     \mathbf{0}_{d_2\times1}&\mathbf{0}_{d_2\times1}&\ldots&\mathbf{0}_{d_2\times1}\\
     \end{array}
     \right],
     \end{equation}
     where the $\star$ {\sa captures}  a zero or a nonzero {\sa entry}. Consequently, the rank of the controllability matrix is equal or less than {\sa $d_1$}. {\mzr Also note that  the controllability matrix  includes  a zero matrix of dimension {\sa $d_2$} by $n$. This contradicts with the earlier assumption.}

\section{Structural controllability {\sa of multi-agent systems under multiple leaders}
\label{sec: Structural controllability of multi-input systems}}
The previous section introduced the necessary and sufficient conditions for the structural controllability {\sa {\mzrg of interconnected agents} under {\blue a} solo leader}. {\mb This result enables us to investigate the structural controllability of {\mkmrr multi-agent} systems with more than one leader. {\sa To this end}, we {\sa first} need to {\sa present} the notion of {\sa \textit{leader-follower connectivity}}.}
\begin{definition}
	The graph representation of {\sa a set of connected agents} is called {\sa \textit{leader-follower connected}}  if there exists at least a leader in each of the {\sa associated} subgraphs which are totally separated from each other.
	\label{def1} 
\end{definition}
 \begin{remark}
 	{\sa There exists an {\mzrg analogy} between the two notion of {\mzr leader-follower} connectivity and having {\blue a} spanning forest rooted at leaders' vertices. Based on the definition, if the graph has a spanning forest rooted at some special vertices, e.g., leaders, there exist at least a path between each node of the graph, except leaders, to one of the leaders' nodes. This property guarantees the existence of at least one leader in every totally separated {\mzrg subgraphs  which}  coincides with Definition \ref{def1}.}

 \end{remark}

 Before presenting the {\sa main} results, it is beneficial to review {\sa some} information about the {\sa whole interconnected} system. Provided
 the multi-agent system has $l$ leaders, the system matrices $A$ and $B$ are of dimensions $(N-l)\times (N-l)$ and $(N-l)\times l${\mkmrr,} {\sa accordingly. Moreover, as} it is mentioned before, there exist{\mkmrr s}  at least one path to each node from one of the leaders in a leader-follower connected topology. {\sa Thus, one can easily conclude that there {\mzr  exist} at least $N-l$ path between leaders and followers.} {\sa The following theorem establishes that the} leader-follower connectivity {
 \sa of {\mzr the  topology} associated {\mzrg with the} graph}  is the necessary and sufficient condition {\sa for the whole interconnected system to be structurally controllable under multiple leaders}.

\begin{thrm}
		{\mzr Consider the} system (\ref{system}) under the communication topology $\mathcal{G}$ {\sa {\mzr with} multiple leaders, i.e., $l>1$}. {\mzrg This system}  is structurally controllable if and only if the system is leader-follower connected.
\end{thrm}
\vspace{-4mm}
\subsection*{Proof of  {\blue Sufficiency}}
The goal is to prove the system is structurally controllable provided that it is leader-follower connected. As mentioned before, a leader-follower connected system with $N$ nodes and $l$ leader{\mkmrr s} has at least $N-l$ {\sa edges}. Based on the results derived in single leader case, {\mkmb if we have exactly $N-l$ edges,} the parameters $c_i$s are linearly independent of each other for every $i\in \mathbf{q}${\sa , where $\mathbf{q}$ is the set of algebraic independent parameters}.  The vectors  $\left[\begin{array}{cc}
     r_{i1}&r_{i2}
     \end{array}\right]$  are independent of each other as well. Now, if the system is leader-follower connected, {\sa again all $c_i$s are linearly independent of each other. {\mzr Hence, it is easy to }see that for every $\mathbf{s}\subset \mathbf{q}$ the matrices $C_{\mathbf{s}}$ and $R_{\mathbf{q}-\mathbf{s}}$ are full rank, namely  $\textnormal{rank}C_\mathbf{s}=f$ and {\mzr  $\textnormal{rank}R_{\mathbf{q}-\mathbf{s}}=\sigma-f$.} Given that we have $N-l$ independent $w_{k}$ to assign, we can conclude {\mzr that $\min_{\mathbf{s} \subset q}(\textnormal{rank}C_\mathbf{s}+ \textnormal{rank}R_{\mathbf{q}-\mathbf{s}})=\sigma=N-l$.} The latter is equal to number of system states.} {\mkmb Besides, as we stated before for the solo leader case, if we have more than $N-l$ edges, it is possible to introduce a subgraph with $N-l$ edges which satisfies the rank condition in (\ref{eq:rankcondition}).} {\sa On the other hand, the term leader-follower connectivity suggests that the system has a spanning forest rooted at the leaders vertices. Due to Proposition \ref{p1}, this means that the corresponding graph of the system is irreducible. Therefore, based on Lemma \ref{l2}, we can {\mzrg conclude that } the transfer graph $\mathcal{T}$ has a spanning tree rooted at ${\mkmr \gamma_{\sigma+1}}$.} 
Based on these two results, the system is structurally controllable.
\vspace{-4mm}
\subsection*{Proof of Necessity}
{\mkm We }{\sa use the proof by contradiction to establish the sufficiency part.  } {\mzr We} assume that the system {\sa was} not leader-follower connected while it {\sa was} structurally controllable. Without the loss of generality, we consider that the system consists of two subsystems which are completely separated from each other. {\sa One of the subsystems is leader-follower connected and includes all the leaders. This subsystem has $N_1$ nodes. The remaining $N_2$ {\mzr nodes can be seen as a second subsystem.}
	{\mkm If we compute the Kalman's controllability matrix for this system, it is easy to show that the controllability rank is equal or less than $N_1$ and the system is not controllable.} 
	 {\mzr This contradicts with the initial assumption and the proof is finished.} 
\section{Conclusion
\label{sec: Conclusion}}
In this paper, the structural controllability of multi-agent systems {\sa under multiple leaders} with fixed topology {\sa was} scrutinized. The necessary and sufficient condition of structural controllability of multi-agent systems {\sa for the both {\mzrg cases} of single and multiple leaders was developed with the help of  the linear parameterization technique}. {\sa We established} that the connectivity {\sa of graph topology}, in {\mzr the }single leader situation, and the leader-follower connectivity {\sa of the associated graph}, in the multi leader case, stand not only as the necessary condition but also as {\blue the} sufficient condition. Some possible future research directions include  investigation of structural controllability condition for switching and linear time-variant {\sa topologies}. }
\ifCLASSOPTIONcaptionsoff
  \newpage
\fi

\bibliographystyle{unsrt}
\bibliography{Ref.bib}

%




\end{document}